\documentstyle[aps,twocolumn,epsfig]{revtex}
\begin{document}
\newcommand{\ve}[1]{\mbox{\boldmath $#1$}}
\twocolumn[\hsize\textwidth\columnwidth\hsize
\csname@twocolumnfalse%
\endcsname
 
\draft

\title {Weakly Interacting Bose-Einstein Condensates Under Rotation:
        Mean-field versus Exact Solutions} 
\author{A. D. Jackson$^1$, G. M. Kavoulakis$^2$, B. Mottelson$^2$, and 
S. M. Reimann$^3$}
\date{\today} 
\address{$^1$Niels Bohr Institute, Blegdamsvej 17, DK-2100 Copenhagen \O,
        Denmark, \\
         $^2$NORDITA, Blegdamsvej 17, DK-2100 Copenhagen \O, Denmark, \\
         $^3$Department of Physics, University of Jyv\"askyl\"a, P.O. Box 35,
         40351 Jyv\"askyl\"a, Finland}
\maketitle
 
\begin{abstract}

   We consider a weakly-interacting, harmonically-trapped Bose-Einstein 
condensed gas under rotation and investigate the connection between
the energies obtained from mean-field calculations and from exact
diagonalizations in a subspace of degenerate states. From the latter
we derive an approximation scheme valid in the thermodynamic limit
of many particles. Mean-field results are shown to emerge as the
correct leading-order approximation to exact calculations in the
same subspace.

\end{abstract}
 
\vskip0.5pc]

Following the observation of Bose-Einstein condensation 
\cite{GSS} in vapours of alkali atoms \cite{RMP}, considerable attention 
has been devoted to the behaviour of these systems under rotation.  
Matthews {\it et al.} \cite{JILA} reported the observation 
of vortex states in a two-component system, and Madison {\it et al.} 
\cite{Madison} demonstrated the existence of vortex states
in a stirred one-component Bose-Einstein condensate.  Numerous
authors have considered the problem of rotation in these systems
theoretically \cite{Rokhsarv,Wilkin,Rokhsar,Fetter,Ben,Bertsch,Feder}.
Many of them focused on the Thomas-Fermi
regime of strong interactions between the atoms 
\cite{Rokhsarv,Fetter,Feder}, where the coherence length is
much smaller than the spatial extent of the system and where the
cloud is expected to exhibit properties like those of a bulk 
superfluid such as $^4$He.  The other interesting limit, to be considered 
here, is that of weak interactions between the atoms, where the coherence
length is much larger than the size of the cloud of atoms and where 
one expects to see properties similar to those of superfluid nuclei
\cite{GC}.

The problem of rotation in the limit of weak interactions has
also been considered previously.  Wilkin {\it et al.} \cite{Wilkin} 
studied the case of effective attraction between the atoms
and showed that, in the ground state of the system, angular 
momentum is carried by the center of mass.  Butts and Rokhsar \cite{Rokhsar}
examined the problem of effective repulsion between the atoms. They 
included interaction effects in mean-field approximation using the 
Gross-Pitaevskii equation and performed numerical calculations of the 
moment of inertia of the system.  In Ref.\,\cite{Ben} one of us 
demonstrated that these systems exhibit two additional kinds of condensation 
associated with the degrees of freedom which come from the multiplicity of 
the degenerate rotating states. Finally, Bertsch and Papenbrock \cite{Bertsch} 
solved the problem exactly within a subspace of degenerate states for
a limited number of particles and angular momenta. 

In the absence of interactions, there is a large degeneracy corresponding to 
the many ways of distributing a given angular momentum among the 
particles \cite{Abra}.  The challenge is to incorporate interactions in 
order to lift the degeneracy.  This can be done either within the mean-field 
approximation \cite{Rokhsar,KMP} or ``exactly'' by diagonalizing the 
Hamiltonian in an appropriate subspace of the degenerate states. In this 
Letter we establish a connection between these two approaches.

Our starting point is the hamiltonian $\hat H=\sum_i \hat h_i + \hat V$, 
where
\begin{eqnarray}
    \hat h_i = - \frac {\hbar^{2}} {2M} {\ve \nabla}_{i}^{2} +
   \frac 1 2 \, M \omega^{2} r_{i}^{2}
\label{h0}
\end{eqnarray}
includes the kinetic energy of the particles and the potential energy
due to the trapping potential and where  
\begin{eqnarray}
   \hat V = \frac 1 2 U_{0} \sum_{i \neq j} \delta({\bf r}_{i} - {\bf r}_{j})
\label{v}
\end{eqnarray}
is the two-body interaction between particles, which is assumed to be of 
zero range.  Here $M$ is the atomic mass, $\omega$ is the
frequency of an isotropic trapping potential, 
and $U_0 = 4 \pi \hbar^2 a/M$ is the strength of the effective two-body 
interaction, with $a$ being the scattering length for atom-atom collisions.
We assume that $a > 0$, i.e., we treat only the case of repulsive 
effective interactions between the atoms.  We further restrict our attention 
to the limit of weak interactions, 
\begin{eqnarray}
   n U_0 \ll \hbar \omega,
\label{wi}
\end{eqnarray}
where $n$ is the density of the condensed atoms.  If the system has angular 
momentum $L$, this condition allows us to work in subspaces of 
states which are degenerate in the absence of interactions.  All other 
states differ by an energy of order $\hbar \omega$, which is
much larger than $n U_0$ by assumption. Finally, we assume that
the cloud of atoms rotates about some axis, and that
the system is in its ground state with respect to this axis, which implies
that our problem is essentially two-dimensional. In the absence of
interactions for the harmonic oscillator potential in two dimensions
the single particle energies are
\begin{eqnarray}
  \epsilon = (2 n_r + |m| + 1) \hbar \omega,
\label{energy}
\end{eqnarray}
where $n_r$ is the radial quantum number, and $m$ is the quantum number
corresponding to the angular momentum. In the lowest energy state of
the system all the atoms are in states with $n_r = 0$,
and with $m$ being zero or having the same sign as the total angular
momentum. 

  Let us describe briefly the two methods which have been used to
attack the problem of rotation, starting with the mean-field approach
\cite{Rokhsar,KMP}. Within this approximation the many-body wavefunction
with $N$ particles and $L$ units of angular momentum $\Psi_{L,N}({\bf r}_1,
{\bf r}_2,\ldots,{\bf r}_N)$ is a Fock state simply expressed as a product
of single particle states:
\begin{eqnarray}
    \Psi_{L,N}({\bf r}_1,{\bf r}_2,\ldots,{\bf r}_N) =
  \Psi({\bf r}_1) \times \Psi({\bf r}_2) \ldots \Psi({\bf r}_N).
\label{nwf}
\end{eqnarray}
It is natural to expand the single-particle states $\Psi({\bf r}_i)$
in the basis of the harmonic-oscillator eigenstates $\Phi_{m}({\bf r}_i)$
with angular momentum $m \hbar$ along the axis of rotation:
\begin{eqnarray}
   \Psi({\bf r}_i) = \sum_{m=0}^{\infty} c_{m} \Phi_{m}({\bf r}_i),
\label{exp}
\end{eqnarray}
where $c_{m}$ are variational parameters. 
The quantity $|c_{m}|^2$ gives the probability 
for the occupancy of state $\Phi_m$. One imposes two constraints on the
parameters $c_{m}$.  The wavefunction must be normalized, 
\begin{eqnarray}
  \sum_{m} |c_{m}^{2}| = 1,
\label{cond1}
\end{eqnarray}
and the expectation value of the angular momentum per
particle must be fixed,
\begin{eqnarray}
  \sum_{m} m |c_{m}|^{2} = L/N.
\label{cond2}
\end{eqnarray}
The expectation value of the interaction energy $\hat V$ 
in the state given by Eq.\,(\ref{nwf}) is
\begin{eqnarray}
   \langle \hat V \rangle = \frac 1 2 N (N-1) U_{0}
\int |\Psi|^{4} \, d{\bf r}.
\label{e}
\end{eqnarray} 
Minimizing this energy with respect to the $c_m$ subject to the 
constraints of Eqs.\,(\ref{cond1}) and (\ref{cond2}), one obtains 
the mean field energy and corresponding wavefunction.

Turning to the ``exact'' approach to this problem \cite{Wilkin,Bertsch}, it 
is convenient to work in the basis $|N_0,N_1,N_2,... \rangle$
where $N_m$ is the number of particles with angular momentum $\hbar m$.
The constraints on particle number and angular momentum now take the form:
\begin{eqnarray}
   \sum_m N_m = N; \, \sum_m m N_m = L.
\label{restr}
\end{eqnarray}
It is convenient to write the interaction energy as
\begin{eqnarray}
   \hat V = \frac  1 2 U_0 \sum_{i,j,k,l} 
   I_{i,j,k,l}  \, a_i^\dagger a_j^\dagger a_k a_l,
\label{intersq}
\end{eqnarray}
where $a_k$ and $a_k^\dagger$ are annihilation and creation operators
respectively, and 
\begin{eqnarray}
   I_{i,j,k,l} &=& 
 \int \Phi_i^*({\bf r}) \Phi_j^*({\bf r}) \Phi_k({\bf r})
  \Phi_l({\bf r}) \, d{\bf r}   \phantom{XXX}
\nonumber \\ \phantom{XXX}
  &=& \frac {(i+j)!} {2^{(i+j)}
 \sqrt{i!\, j! \, k! \, l!}} \int |\Phi_0({\bf r})|^4 d{\bf r}
\label{integral}
\end{eqnarray}
when $i+j=k+l$ and zero otherwise.  The final step is to diagonalize the 
matrix $\hat V$ in the space of the states $|N_0,N_1,N_2,... \rangle$ 
which satisfy the constraints of Eq.\,(\ref{restr}).

Before turning to a comparison of these methods, it is useful to consider 
their advantages and disadvantages.  On the one hand, the mean-field 
approximation gives the energy in the asymptotic limit $L$ and $N 
\rightarrow \infty$ with $L/N$ fixed.  This limit is not directly accessible 
to the exact method since the dimensionality of the space increases 
dramatically with increasing $L$ and $N$.  On the other hand, the exact 
method provides the entire spectrum of excited states and a lower ground 
state energy than the mean-field result.  The mean-field assumption of 
a simple product wavefunction ignores correlations between the particles
which are contained in the exact diagonalization.  The mean-field energy 
offers only an upper bound for the true ground state energy.  To be more 
specific, the leading term in the ground state energy of the system must 
grow like $N^2$ for the interaction considered.  As we show below, the 
energy obtained by the exact method in the asymptotic limit is bounded from 
below by an expression which has the same $N^2$ coefficient as the
mean-field result.  Thus, the exact energy differs from the mean-field 
value by terms of order $N$ or less.   In addition, we will suggest 
simple methods for computation of the order $N$ correction to the ground 
state energy in the asymptotic limit. 

The spirit of the mean field approximation is that the wavefunction is 
composed of a large number of states with occupation numbers in the vicinity 
of $N |c_m |^2$ but that these states are of measure zero in the full Hilbert 
space considered.  Up to a staggering related to the phases of the $c_m$, 
the wavefunction should then be a smooth and differentiable function of 
occupation number for single particle states containing a finite fraction of 
the particles.  Such states are amenable to simple treatment.  States with 
$c_m = 0$ in a mean field description will require special treatment.  
Consider the particular case $L=2N$ using a Hilbert space $0 \le m \le 4$.  
In mean field theory it is found that the only non-zero coefficients are $c_0$,
$c_2$, and $c_4$. Initially retaining only these states, we will consider the 
basis of states  
\begin{eqnarray}
   | k \rangle  = |k, 0, N-2k, 0, k \rangle.
\label{l2n}
\end{eqnarray}
Write the wavefunction as 
\begin{eqnarray}
   |L=2N, N \rangle = \sum_k (-1)^k \psi_k
 \, |k \rangle, 
\label{l2ngs}
\end{eqnarray}
where the $k$-dependent phase has been chosen to minimize the resulting 
energy.   The eigenvalue equation has the form
\begin{eqnarray}
  \hat V_{k,k} \psi_k - \hat V_{k,k-1} \psi_{k-1} - \hat V_{k,k+1} \psi_{k+1}
 = {\cal E}_{2N, N} \psi_k,
\label{eigeq}
\end{eqnarray}
where $\hat V_{k,k'} = \langle k | \hat V | k' \rangle$ is the matrix
element of the interaction between the states $|k\rangle$ and 
$|k' \rangle$ given by Eq.\,(\ref{l2n}).  The two-body interaction
$\hat V$ connects only states for which $k - k'=\pm 1$, or 0. 

Assuming that $\psi_k$ is smooth and differentiable over a suitable range, 
we expand the wavefunction as 
\begin{eqnarray}
  \psi_{k \pm 1} = \psi_k \pm \partial_k \psi_k 
 + \frac 1 2 \partial^2_k \psi_k
\label{expansion}
\end{eqnarray}
to obtain the eigenvalue equation  
\begin{eqnarray}
   -\frac 1 2 (\hat V_{k,k-1} + \hat V_{k,k+1}) \, \partial^2_k \psi_k 
 + V_{\rm eff} \psi_k = {\cal E}_{2N, N} \psi_k,
\label{eigeqhp}
\end{eqnarray}
where the effective potential is
\begin{eqnarray}
  V_{\rm eff}(k) = \hat V_{k,k} - \hat V_{k,k-1} - \hat V_{k,k+1}.
\label{lead}
\end{eqnarray}
The leading term in the energy is of order $N^2$ and is simply the minimum 
value of $V_{\rm eff}(k)$.  This minimum occurs at 
\begin{eqnarray}
  k_0 = N \frac {16 \sqrt 6 - 28} {64 \sqrt 6 - 109}.
\label{k0}
\end{eqnarray}
The occupancy of the $m=0$ and $m=4$ states is $k_0 / N$ in the
asymptotic limit; the occupancy of $m=2$ is $1 - 2 k_0/N$.  This is 
precisely the result found in Ref.\,\cite{KMP} within the mean-field 
approximation.  Numerical exploration for $N > 10^9$ indicates that 
the energy of the minimum of the effective potential, $V_{\rm eff}(k_0)$, 
can be described by the approximate form
\begin{equation}
   \tilde {\cal E}_{2N,N} \approx 
\left( 0.177256895 N^2 - 0.366043 N + {\cal O}( N^0 ) \right) v_0,
\label{l2ngse}
\end{equation}
where $v_0 = U_0 \int |\Phi_0|^4 \, d{\bf r}$, with $\Phi_0({\bf r})$
being the ground state of the harmonic oscillator.
As expected, the $N^2$ coefficient in this expression is identical to 
the corresponding term in the mean field \cite{KMP}.  In general, 
the properties of the minimum of $V_{\rm eff}(k)$ reproduce the leading 
$N^2$ term of mean field theory.  

Expanding $V_{\rm eff}(k)$ around $k_0$, we find that the effective potential 
has the form 
\begin{eqnarray}
  V_{\rm eff}(k) = \tilde {\cal E}_{2N,N} + \frac 1 2 d \, (k - k_0)^2 v_0,
\label{leadexp}
\end{eqnarray}
where $\tilde {\cal E}_{2N,N}$ is given by Eq.\,(\ref{l2ngse}) and 
where the constant $d \approx 0.3732$ is necessarily positive and of 
${\cal O}(N^0 )$.  The leading, ``kinetic energy'' term in 
Eq.\,(\ref{eigeqhp}) is positive with $\hat V_{k,k-1} + \hat V_{k,k+1} 
\approx 0.0381 N^2$.  That this term is of ${\cal O}( N^2 )$ is a consequence 
of the fact that ${\hat V}$ is a two-body operator.  The ground-state energy 
for this harmonic oscillator problem is readily found to be  
\begin{eqnarray}
  {\cal E}_{2N,N} = \tilde {\cal E}_{2N,N} +  \frac 1 2 \hbar \Omega
\label{ho}
\end{eqnarray}
with 
\begin{eqnarray}
\hbar \Omega = \sqrt {d (\hat V_{k,k-1} + \hat V_{k,k+1})} \approx 
0.1192 N v_0.
\label{hq}
\end{eqnarray}
The effective oscillator length corresponding to Eq.\,(\ref{eigeqhp}) is 
proportional to $([{\hat V}_{k,k-1} + {\hat V}_{k,k+1}]/d)^{1/4}$ and is of 
order $N^{1/2}$.  This demonstrates the consistency of our assumptions:  
The ground state wavefunction $\psi_k$ is a Gaussian centered at $k_0$ with 
a width of order $N^{1/2}$.  It contains significant contributions
from a large number of states, which, however, is only a
vanishingly small fraction of all the states in our Hilbert space.  Combining 
Eqs.\,(\ref{l2ngse}), (\ref{ho}), and (\ref{hq}) we obtain a ground state 
energy of 
\begin{equation}
{\cal E}_{2N,N} = \left(  0.177256895 N^2 - 0.306406 N + {\cal O}( N^0 ) 
\right) v_0.
\label{hof}
\end{equation}

   The general conclusion from this analysis is that the correct 
${\cal O}( N^2 )$ contribution to the energy can be obtained 
by determining the classical minimum of the effective potential energy, 
$V_{\rm eff}(k)$.  Quantum fluctuations lead to corrections of order $N$.  
Since the kinetic energy term is positive, these corrections are necessarily 
positive and this leading term thus provides a lower bound on the ground 
state energy.  By contrast, the mean-field approach leads to  
minimizing the same classical potential but provides an 
upper bound on the energy as a consequence of its variational character.  
Thus, the ground state energy is trapped between upper and lower bounds with 
the same leading term.  Thus, we conclude that the ${\cal O}( N^2 )$ 
contribution to the ground state energy obtained by mean field theory is 
exact in the large $N$ limit.  

While the space considered in the above example was severely limited, 
this method can be generalized to include as many states as necessary.  
Our analysis holds provided that the states included are occupied by 
a finite fraction of the total number of particles.  If this is true, 
roughly $N^{1/2}$ states for each of the variables in the 
multi-dimensional space will make significant contributions
to the wave function and the approximation of
differentiability is valid.  One can then derive an eigenvalue equation
of the form of Eq.\,(\ref{eigeqhp}), of multi-dimensional coupled 
harmonic oscillators.  A crucial assumption in establishing the 
asymptotic validity of the mean field result is the positivity of the 
``inertial'' parameters such as the coefficient of $\partial_k^2 \psi_k$ in 
the case of Eq.\,(\ref{eigeqhp}).  We have verified that this is true
in a number of specific cases, and we believe that it is 
true in general.

Let us now return to the states $m=1$ and $m=3$ in our example.  These 
states are not macroscopically occupied, i.e., the corresponding $c_1$ 
and $c_3$ are zero in a mean field calculation.  The expansion of 
Eq.\,(\ref{expansion}) cannot be used for their description, and their 
incorporation requires some care.  To include their effects, we 
approximate $| L = 2N , N \rangle_0$ by the single state $|k = k_0 \rangle$ 
of Eq.\,(\ref{l2n}).  Starting with this state, we build a basis of states 
containing $n_1$ particles with $m=1$ state and $n_3$ particles with $m=3$, 
being careful to conserve particle number and angular momentum. The effect
of these states with odd $m$ values can then be obtained by diagonalizing 
${\hat V}$ in this basis.  If $0 \le (n_1 + n_3 ) \le 2Q$, there will be 
$(Q+1)^2$ states in this basis.  Convergence of this calculation to 
arbitrary accuracy can be obtained by choosing $Q$ to be large but 
none the less of order $N^0$.  This confirms the fact that the states 
$m=1$ and $m=3$ are not macroscopically occupied.  In the present case, the 
lowest eigenvalue of this matrix converges rapidly to an asymptotic 
value of $-0.1812 N v_0$.  The linear contribution to the energy from such 
states is general.  An improved description of $| L = 2N , N \rangle_0$ 
would lead to changes in this energy of order $N^0$.
 
We are now able to obtain the asymptotic ground state energy for the case 
$L = 2N$ correct to order $N$ by combining this correction for the odd $m$ 
states with the result of Eq.\,(\ref{hof}).  We find 
\begin{eqnarray}
   {\cal E}_{2N,N} \approx \left( 0.177256895 N^2 - 0.487576 N + 
{\cal O}( N^0 ) 
\right) v_0.
\label{pertdv}
\end{eqnarray}
This result is in remarkable agreement with the expression
\begin{eqnarray}
{\cal E}_{2N,N} \approx \left( 0.177256752 N^2 - 0.487848 N + 
{\cal O}( N^0 ) \right) v_0 
\label{stepnuml2n}
\end{eqnarray}
obtained by direct numerical diagonalization in the same truncated
space $0 \le m \le 4$ for values of $100 \le N \le 220$.  Note that 
the case $N = 220$ requires diagonalization of a matrix of dimension 
$305990$.  The small discrepancies obtained are a consequence of the 
limited range of $N$ for which exact diagonalization is possible and 
the largest values of $Q$ considered in our approximate calculation.  These 
restrictions are absent in calculations within the same Hilbert space 
$0 \le m \le 4$ for $L = N/2$.  It is then possible to consider 
$N$ as large as $600$.  All states have an occupation number of order $N$, 
and the final matrix diagonalization in our approximate treatment is not 
required.  In this case, the quadratic terms in the ground state energy 
agree to $10$ significant figures, and the linear terms differ by $1$ part 
in $10^5$.  

Additional evidence comes from exact diagonalizations (with no truncation in 
$m$) for the case $L=2N$ for $4 \le N \le 24$.  The resulting energies are 
well described by 
\begin{eqnarray}
   {\cal E}_{2N, N} \approx (0.1749 N^2 - 0.5206 N + 0.0028) v_0.
\label{entl2n}
\end{eqnarray}
As argued above, mean field theory suggests that only states with even $m$ 
are macroscopically occupied in the ground state for $L=2N$.  This is 
supported by exact diagonalizations for the same range of $N$ from which we 
have excluded single-particle states with odd $m$.  The energy is then 
found to be 
\begin{eqnarray}
   {\cal E}_{2N, N} \approx (0.1755 N^2 - 0.2980 N - 0.0905) v_0.
\label{entl2ntr}
\end{eqnarray}
Comparing the coefficients of $N^2$ and $N$ in Eqs.\,(\ref{entl2n}) 
and (\ref{entl2ntr}), we see that only the latter is materially affected 
by this truncation.  This is consistent with our assertion that states 
which are not occupied macroscopically contribute to the energy 
at order $N$. 

In summary we have adopted a ``correspondence principle approach'' to describe 
repulsive two-body interactions between the particles in a Bose-Einstein 
condensate by assuming that the wavefunction in the $N \to \infty$ limit is 
a smooth and differentiable function of the occupation number.  This 
approach leads us to the result that mean field theory provides an 
exact determination of the $N^2$ term in the ground state energy and permits 
determination of the contribution to this energy of order $N$. Butts and
Rokhsar have arrived at a similar conclusion \cite{Rokhsar}. Evidently, 
the same arguments can be used to describe the spectrum of excited states.  
The present methods would appear to be a useful and practical supplement 
to mean field methods whenever these are of value.

\vskip1pc

   G.M.K. was supported by the European Commission, TMR program, under
contract No.\,ERBFMBICT 983142, and S.M.R. under contract 
No.\,ERB4001GT 970292. We would like to thank C. J. Pethick for
helpful discussions and M. Koskinen for advice regarding numerical
questions. G.M.K. would like to thank the Foundation of Research and 
Technology, Hellas (FORTH) for its hospitality, and S.M.R. would
like to acknowledge financial support from the ``Bayerische Staatsministerium
f\"ur Wissenschaft, Forschung und Kunst".


\begin{references}

   \bibitem{GSS} A. Griffin, D. W. Snoke, and S. Stringari (Cambridge
   University Press, Cambridge, England, 1995).

   \bibitem{RMP} F. Dalfovo, S. Giorgini, L. P. Pitaevskii, and S. Stringari, 
    Rev. Mod. Phys. {\bf 71}, 463 (1999). 

   \bibitem{JILA} M. R. Matthews, B. P. Anderson, P. C. Haljan, D. S. Hall,
   C. E.Wieman, and E. A. Cornell, Phys. Rev. Lett. {\bf 83}, 2498 (1999).

   \bibitem{Madison} K. W. Madison, F. Chevy, W. Wohlleben, and J.
    Dalibard, Phys. Rev. Lett. {\bf 84}, 806 (2000).
   
   \bibitem{Rokhsarv} D. S. Rokhsar, Phys. Rev. Lett. {\bf 79}, 2164 (1997).

   \bibitem{Wilkin} N. K. Wilkin, J. M. F. Gunn, and R. A. Smith,
   Phys. Rev. Lett. {\bf 80}, 2265 (1998).

   \bibitem{Rokhsar} D. A. Butts, and D. S. Rokhsar, Nature
   {\bf 397}, 327 (1999).

   \bibitem{Fetter} M. Linn, and A. L. Fetter, e-print cond-mat/9906139.
  
   \bibitem{Ben} B. Mottelson, Phys. Rev. Lett. {\bf 83}, 2695 (1999).

   \bibitem{Bertsch} G. F. Bertsch, and T. Papenbrock, Phys. Rev. Lett. 
{\bf 83}, 5412 (1999).

   \bibitem{Feder} D. L. Feder, C. W. Clark, and B. I. Schneider,
   e-print cond-mat/9910288; e-print cond-mat/9904269.

   \bibitem{GC}  G. Baym and C. J. Pethick, Phys. Rev. Lett. {\bf 76}, 6
   (1996).

  \bibitem{Abra} M. Abramowitz and I. A. Stegun, {\it Handbook of Mathematical
   Functions} (National Bureau of Standards, Washington, DC, 1964), p. 825.
   
  \bibitem{KMP} G. M. Kavoulakis, B. Mottelson, and C. J. Pethick, 
  e-print cond-mat/0004307.

\end{references}
\end{document}